\documentclass[aps,prl,superscriptaddress,floatfix,reprint,showkeys,showpacs]{revtex4-1}

\usepackage{graphicx}
\usepackage{subfigure}
\usepackage{float}
\usepackage[space]{grffile}
\usepackage{leftidx}
\usepackage{amsmath,amssymb,amstext}
\usepackage{xcolor}
\usepackage{wrapfig}

\newcommand{\Tf}{T_\mathrm{f}}
\newcommand{\Ti}{T_\mathrm{i}}
\newcommand{\Veff}{V_\mathrm{eff}}
\def\m1cond{$m_F{=}-1$}
\begin{document}
\title{Condensing magnons in a degenerate ferromagnetic spinor Bose gas}
\author{Fang Fang}
\affiliation{Department of Physics, University of California, Berkeley, California 94720, USA}
\email{akiraff@berkeley.edu}
\author{Ryan Olf}
\affiliation{Department of Physics, University of California, Berkeley, California 94720, USA}
\author{Shun Wu}
\affiliation{Department of Physics, University of California, Berkeley, California 94720, USA}
\author{Holger Kadau}
\affiliation{5. Physikalisches Institut and Center for Integrated Quantum Science and Technology,
Universit\"at Stuttgart, Pfaffenwaldring 57, 70569 Stuttgart, Germany}
\author{Dan M. Stamper-Kurn}
\affiliation{Department of Physics, University of California, Berkeley, California 94720, USA}
\affiliation{Materials Sciences Division, Lawrence Berkeley National Laboratory, Berkeley, California 94720, USA}

\begin{abstract}
We observe the condensation of magnon excitations within an $F=1$ $^{87}$Rb spinor Bose-Einstein condensed gas.  Magnons are pumped into a longitudinally spin-polarized gas, allowed to equilibrate to a non-degenerate distribution, and then cooled evaporatively at near-constant net longitudinal magnetization whereupon they condense.  We find magnon condensation to be described quantitatively as the condensation of free particles in an effective potential that is uniform within the ferromagnetic condensate volume, evidenced by the number and distribution of magnons at the condensation transition.  Transverse magnetization images reveal directly the spontaneous, inhomogeneous symmetry breaking by the magnon quasi-condensate, including signatures of Mermin-Ho spin textures that appear as phase singularities in the magnon condensate wavefunction.
\end{abstract}
\maketitle

Magnons are collective spin excitations of a magnetically ordered medium.  At thermal equilibrium, these bosonic quasiparticles are not conserved in number, and, thus, are not expected
to undergo Bose-Einstein condensation.  Nevertheless, magnon condensation has been observed in non-equilibrium systems that are pumped with an excess population of magnons that decays
slowly \cite{volo0820years,demo06magnonpumping,vain15magnon}.  Magnon condensation also describes the onset of transverse canted magnetic order in spin-dimer compounds at high
magnetic fields \cite{giam99ladder,niku00magnon}.  Signatures of magnon condensation include properties of the critical point \cite{niku00magnon,seba06reduction}, the
accumulation of magnons in low-energy states \cite{demo06magnonpumping}, and spontaneous symmetry breaking indicated by the emergence of large regions of precessing transverse
magnetization \cite{volo0820years,vain15magnon,demi08spont}.


Here, we report on magnon condensation in a ferromagnetic superfluid gas.  We use ultracold spinor Bose-Einstein gases of $^{87}$Rb, for which the longitudinal magnetization is nearly a conserved quantity \cite{stam13rmp}.  Thus, we are able to produce a long-lived excess of magnon excitations above the critical number for magnon condensation.  Like in other
systems, we detect magnon condensation by determining the critical point, observing the accumulation of magnons in low-energy states, and probing for the spontaneous transverse
magnetization of the atomic gas.  Compared with other systems exhibiting magnon condensation, the atomic spinor gas is described by a simple theoretical model that we compare
quantitatively to the measured magnon distribution at the phase transition.

A key finding of our work is that, at the limit of low temperature, magnon condensation occurs as the Bose-Einstein condensation of free particles in a uniform potential.  Within an
inhomogeneous trapping potential, a ferromagnetic spinor Bose-Einstein condensate equilibrates at a non-uniform density and a uniform chemical potential.  In the absence of
spin-dependent potentials and dipolar interactions \cite{mart14magnon}, rotational symmetry implies that magnon excitations of the ferromagnetic superfluid are gapless, so that their
effective potential is uniform within the condensate volume.  Like free particles, the magnons disperse quadratically with wavevector, with an effective mass that is, within a weakly
interacting gas, nearly equal to that of a bare atom \cite{phuc13,mart14magnon}.  A box-like trap for atoms in a single spin state was recently constructed using a finely tuned optical
and magnetic trap \cite{gaun13flat}.  In contrast, the box-like potential for magnon excitations is produced naturally by the gas itself, reducing sensitivity to experimental
imperfections.  Such potentials yield quantum gases at nearly uniform density, allowing for precise measurements of system properties that would otherwise be smeared out by
inhomogeneous broadening. The condensation of minority spins in a spinor Bose gas was previously observed at high temperature where interactions play little role in modifying the
inhomogeneous trapping potential \cite{erha04constant}.

Our experiments begin with a Bose-Einstein condensed $^{87}$Rb gas, fully magnetized in the $|F=1, m_F = -1\rangle$ hyperfine state, and held in a state-independent optical trap.  The
gas is exposed to a 177 mG magnetic field that is uniform to about 10 $\mu$G within the condensate.  The gas temperature is controlled through evaporative cooling by the depth of the
optical trap. With the gas at an initial temperature ranging from $\Ti \simeq 80 \, \mbox{nK}$ to $140 \, \mbox{nK}$, we ``pump'' magnons into the gas by applying a spatially uniform, pulsed rf magnetic field at the 124
kHz Larmor precession frequency.  The pulse tips the atomic spin by a slight angle (up to 0.7 rad) at which the population of spin-flipped atoms is dominantly in the $|F=1, m_F =
0\rangle$ state.  The gas is then allowed to thermalize at constant magnetization and at constant trap depth for $2.5\,\mbox{s}$. By measuring its momentum distribution, we
confirm that the magnon gas at this initial stage is not condensed.

We then cool this magnon-imbued gas by lowering the optical trap depth over a time $t_\mathrm{ramp}=5\,\mbox{s}$ and then maintain a constant depth for
$t_\mathrm{hold}=2.5\,\mbox{s}$ so that the gas reaches a steady state temperature $\Tf$ before the gas is probed. During these times, not only does the total number of trapped atoms
drop, but also the fractional population of magnon excitations drops owing to the preferential evaporation of magnons from the trap \cite{olf15cooling}.  Nevertheless, provided that the
initial magnon number is high enough, the magnon population at the time of probing can be sufficient to cross the magnon condensation transition.  The coldest samples studied in this
work are prepared at $\Tf = 30$ nK in a trap with depth $k_B \times 122$ nK and trap frequencies $\omega_{x,y,z}=2\pi \{23, 8.8, 160\} \, \mbox{s}^{-1}$, with final magnon populations
varying between 4\% and 10\% of the $6.5 \times 10^5$ atoms in the $m_F = -1$ condensate.

\begin{figure}[t]
	\begin{center}
		\includegraphics[width=0.46\textwidth]{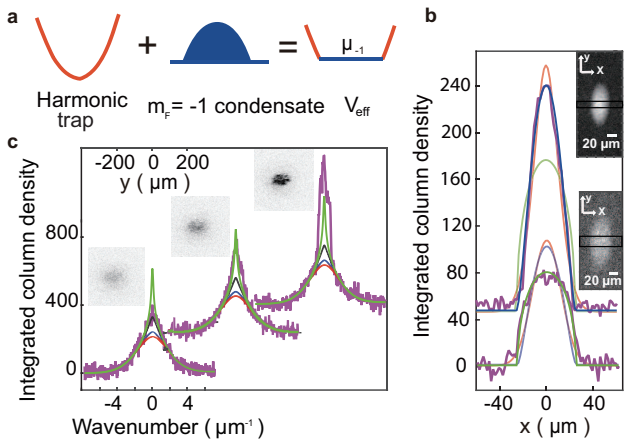}
		\caption{Magnons in a flat potential. (a) The effective potential $V_\mathrm{eff}$ for magnon excitations is the sum of the harmonic trap potential (red) and the repulsive potential from the inhomogeneous ferromagnetic condensate (blue, maximum potential of $\mu_{-1}$), shown schematically in one dimension.  (b) The column density of a low temperature non-degenerate magnon gas (bottom column density image) is large within the volume of the $m_F=-1$ condensate (top column density image).  Line densities (integrated within black rectangles) are fitted to predicted critical density in a harmonic trap (red), constant density within condensate volume (green, best fit for magnon distribution), or Thomas-Fermi condensate distribution (blue, best fit for the $m_F=-1$ condensate). (c) Images and integrated line profiles (purple) of magnon momentum distribution below (left), at (center) and above (right) critical number for magnon condensation.  Profiles are fitted to Bose distributions $\propto g_\alpha(z(p))$ with $z(p)$ defined in the text and $\alpha = \{1, 1.5, 2\}$ (black, blue, red lines, respectively), and the distribution expected in $V_\mathrm{eff}$ (green).}
		\label{fig:Box_distribution}
	\end{center}
\end{figure}

The properties of the magnon gas at the condensation transition are strongly influenced by the effective potential in which the magnons propagate. This effective potential
$V_\mathrm{eff}(\mathbf{r})$ is the sum of the external trapping potential $V(\mathbf{r})$ (with $V=0$ at the trap center) and the interaction energy $V_\mathrm{int}(\mathbf{r})$ of an
$m_F = 0$ atom within the $m_F = -1$ condensate (Fig.\ \ref{fig:Box_distribution}a).  According to mean-field theory, $V_\mathrm{int}(\mathbf{r}) = \mu_{-1} - V(\mathbf{r})$ within the
condensate volume and zero otherwise, with $\mu_{-1}$ the chemical potential of the $m_F =-1$ gas.  We derive this expression using the Thomas-Fermi approximation and noting the
equality $a_{-1,-1} = a_{-1,0}$ implied by rotational symmetry of the contact interaction;  here, $a_{i,j}$ is the s-wave scattering length for a collision between atoms in the $|m_F =
\{i,j\}\rangle$ states.  The effective potential is then $V_\mathrm{eff}(\mathbf{r}) = \max(V(\mathbf{r}), \mu_{-1})$\footnote{Here we neglect effects of non-zero temperature,
magnon-magnon interactions, and spin-dependent potentials including the linear Zeeman energy of the uniform magnetic field, which, neglecting dipolar interactions, can be gauged away by
treating the gas in a rotating frame}.

The first signature of the box-like potential for magnon excitations is the position-space distribution of the non-degenerate magnon gas.  The normal magnon density is expected to be
constant within the volume of the ferromagnetic condensate, and then, at low temperature (defined as $k_B T \ll \mu_{-1}$), to diminish rapidly outside that volume.  We image this
distribution within our coldest samples, for which $k_B \Tf /\mu_{-1}\sim0.7$, by applying a microwave pulse that drives atoms selectively from the $|F=1, m_F = 0\rangle$ state to the
$|F=2\rangle$ hyperfine state, and then imaging selectively the $F=2$ atoms with resonant probe light. The observed column density of the magnon gas $\tilde{n}_0$, shown in Fig.\
\ref{fig:Box_distribution}b, is indeed large within the area defined by the Thomas Fermi radii of the condensate, $R_{x,y} = (2 \mu_{-1} / m \omega^2_{x,y})^{1/2} = \{21, 54\} \, \mu
\mbox{m}$ in the imaged directions, and is described quantitatively by the function $\tilde{n}_0 \propto \max\left(1-x^2 / R_x^2 - y^2 / R_y^2,0\right)^\beta$ with $\beta = 1/2$.  In
contrast, the non-uniform density of the majority-spin condensate leads to a column density with a different exponent, $\beta = 3/2$.  Both distributions deviate from that of a
harmonically trapped critical Bose gas, $\propto g_2( \exp[-x^2/\sigma_x^2 - y^2 / \sigma_y^2])$ with $g_\alpha$ being the polylogarithm function of order $\alpha$ and $\sigma_{x,y} =
(k_B \Tf /m \omega_{x,y}^2)^{1/2} = \{12, 31\} \, \mu \mbox{m}$.

\begin{figure}[t]
	\begin{center}
		\includegraphics[width=0.46\textwidth]{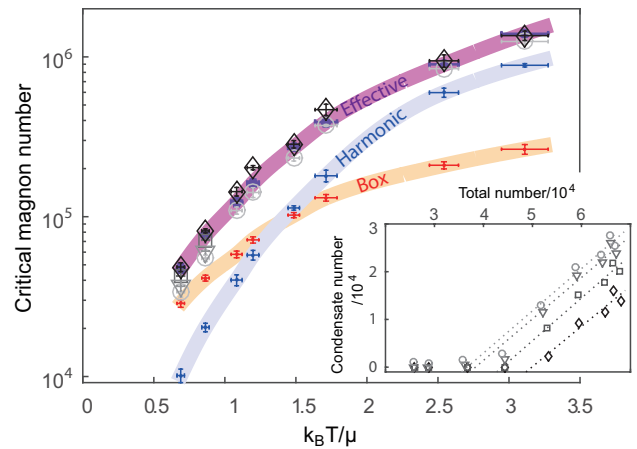}
		\caption{The measured critical magnon number $N_{\mathrm{mag,c}}$ (black markers) is compared to the expected value (blue, red, purple points, respectively) for magnons in a harmonic, box (size defined by condensate volume), and $V_\mathrm{eff}$. Shaded lines are guides to the eye. Error bars on model predictions account for statistical uncertainties in experimental parameters.  Inset: The magnon condensate number, shown for $k_B \Tf/\mu_{-1} = 0.7$, is the integrated residual above the fits to the non-condensed magnon momentum distributions using one of four fit functions: $g_\alpha(z(\boldsymbol{\rho}))$ with $\alpha = 1$ (squares), 1.5 (triangles), or 2 (circles), or the predicted distribution in the effective potential (diamonds).  A linear fit shows the magnon condensate number increasing linearly above $N_\mathrm{mag,c}$; error bars are statistical. Extracted measured critical magnon numbers from the four fitting functions are shown in the main plot (diamonds, squares, triangles, circles with a decrease in grayscale). }
		\label{fig:Ncritical}
	\end{center}
\end{figure}

The second manner in which the box-like potential for magnon condensation is evident is the variation of the critical magnon number with temperature. At each temperature $\Tf$, ranging
from 30 to 114 nK, we examine the momentum distribution of magnons as a function of total magnon number.  This momentum distribution is measured by
releasing all atoms from their trap, and then using state-selective magnetic-field focusing and absorption imaging \cite{olf15cooling}. Excluding data from the central region of the image, we fit the momentum-space column density to several parametrizations of the distribution of a non-condensed Bose gas at variable magnon chemical potential
and temperature $\Tf$.  These parametrizations include $g_\alpha\left(z(\mathbf{p})\right)$ with $\alpha$ chosen among several values \cite{gaun13flat}, which describes the expected momentum distribution of bosons in various power-law potentials, and also the distribution expected for bosons trapped in the effective potential $V_\mathrm{eff}$; here, $z(\mathbf{p}) = \exp\left[ (\mu - p^2/2m)/k_B \Tf\right]$ with $\mu = \mu_0 - \mu_{-1}$ being the magnon chemical potential referenced to $\mu_{-1}$ and ignoring the Zeeman energy ($\mu=0$ at the magnon condensation transition), and $\mathbf{p}$ being the momentum in the imaged plane. The size of the excluded region is chosen such that these functions give about the same temperature.  The magnon condensate number is then determined by subtracting the fitted function from the observed distribution and summing over the image, including the central region. This magnon condensate number rises linearly from zero with increasing total magnon number (Fig.\ \ref{fig:Ncritical}c inset) above a number $N_{\mathrm{mag,c}}$ that we identify as the critical magnon number for condensation.  Different parametrizations of the non-condensed magnon distribution give critical magnon numbers that vary by around 10\% at the same $\Tf$ for our coldest samples.

In Fig.\ \ref{fig:Ncritical}, we compare $N_{\mathrm{mag,c}}$ to predicted values for ideal-gas Bose-Einstein condensation in one of three potentials: a harmonic trapping potential, a
hard-walled box potential with a volume matching that of the $m_F = -1$ condensate, and the effective trapping potential $\Veff$, using the experimentally determined
$\omega_{x,y,z}$, $\mu_{-1}$, and $\Tf$ at each setting \footnote{See Supplemental Material at [URL will be inserted by publisher] for predicted critical magnon number in each
potential.}.  The measurements agree well with predictions based on the condensation of magnons in the effective potential $\Veff$.  At low temperature, the measured critical magnon
number tends toward that predicted for a hard-walled box potential while at high temperatures ($k_B \Tf \gg \mu_{-1}$), $N_{\mathrm{mag,c}}$ tends toward the prediction for a
harmonically trapped gas, as expected.

The details of the critical magnon momentum distribution serve as a third signature of condensation in a box-like potential.  We assume that the calculations based on magnons propagating in
the effective potential give the correct critical magnon number, and use this number to identify the momentum-space image of the magnon gas just below the magnon
condensation transition.  As shown in Fig $\ref{fig:Box_distribution}$b, this distribution is more sharply peaked at low momentum than predicted for atoms expanding from a harmonic
potential.  The data are consistent with the expected momentum-space distribution of magnons in the effective potential, although we do not observe the expected sharp cusp
at zero momentum ($\alpha=1/2$ for a uniform potential), likely owing to limited momentum resolution.

Magnon condensation is a symmetry breaking phase transition.  In our spinor gas system, prior to magnon condensation, the longitudinally magnetized spinor Bose-Einstein condensate
retains the $O(2)$ spin-gauge symmetry that describes the combination of rotations about the longitudinal (magnetic field) axis and the multiplication of the condensate order parameter
by a phase.  Upon magnon condensation, this symmetry is broken, the ferromagnet now acquiring a non-zero transverse magnetization that serves as the order parameter for the magnon
condensate.

We detect this transverse magnetization using absorptive state-selective \emph{in-situ} imaging \cite{mart14magnon}.  We apply an rf pulse to rotate the atomic spins by $\pi/2$ and then
measure the column density in each of the Zeeman sublevels using a sequence of brief microwave and optical probe pulses.  The difference in the $m_F = \pm 1$ column densities gives one
component of the dimensionless transverse column magnetization $\tilde{M}_x$ at the start of the several-ms-long measurement procedure.  We then apply a second, carefully timed $\pi/2$
spin rotation, again measure the Zeeman state distribution, and subtract the $m_F = \pm 1$ images to obtain $\tilde{M}_y$.  We intersperse spin-echo pulses within the measurement
sequence to control for magnetic field variations, similar to Ref.\ \cite{guzm11}.

Assuming the majority spin condensate order parameter $\psi_{-1}(\mathbf{r})$ is known, the magnon condensate wavefunction $\psi_0(\mathbf{r})$ is characterized interferometrically by the transverse magnetization through the relation $M_T(\mathbf{r}) = M_x(\mathbf{r}) + i M_y(\mathbf{r}) = \sqrt{2} \psi_{-1}^*(\mathbf{r}) \psi_0(\mathbf{r})$.  This
relation is valid when the longitudinal magnetization of the gas is sufficiently large.  We measure instead the transverse column magnetization, $\tilde{M}_T$, which can be taken as
$\tilde{M}_T(\boldsymbol{\rho}) = e^{i \varphi(\boldsymbol{\rho})} \sqrt{2 \tilde{n}_{-1,\mathrm{c}}(\boldsymbol{\rho}) \tilde{n}_{0,\mathrm{c}}(\boldsymbol{\rho})}$ if we assume the
atomic spin state of the condensed atoms is constant along the imaging axis.  Here, $\tilde{n}_{m_F, \mathrm{c}}$ is the column density of the $m_F$ component of the condensate,
$\varphi$ is the magnon condensate phase (up to a uniform offset), and $\boldsymbol{\rho}$ is the position in the imaged plane.  We assume here that the majority-spin condensate has
uniform phase given that this condensate is prepared by very gradual evaporative cooling, and that it contains no vortex excitations as confirmed by high-resolution imaging after a
short time of flight from the optical trap.

\begin{figure}[t]
	\begin{center}
		\includegraphics[width=0.45\textwidth]{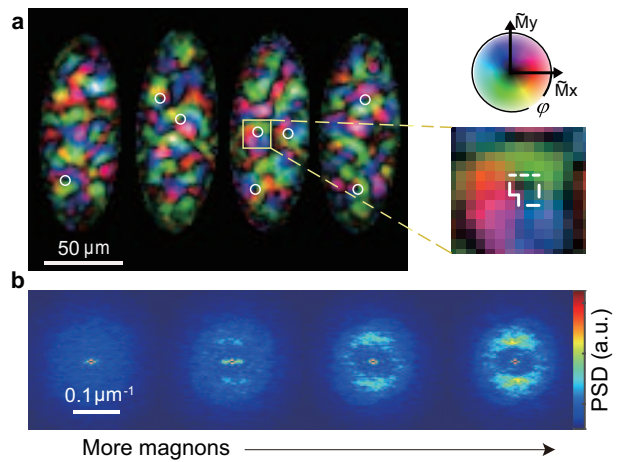}
		\caption{Inhomogeneous transverse magnetization measured for four experimental repetitions, in coldest samples and with $\sim 10\%$ magnon fraction, with hue and brightness representing orientation and amplitude, as indicated by color wheel.  Identified phase singularities are marked by white circles.  The inset highlights one such singularity, with the dashed white line showing a path along which the magnon order parameter phase winds by $2\pi$.  (b) Spatial Fourier power spectral density (PSD) of transverse magnetization averaged over 20 experimental repetitions, for magnon fraction increasing from 7\% (below critical value) to 10\% (above critical value).}
		\label{fig:vortex}
	\end{center}
\end{figure}

The images taken on samples with sufficiently large magnon populations (Fig.\ \ref{fig:vortex}a) show significant transverse magnetization within the boundary of the ferromagnetic
condensate.  The magnon condensate number implied by these images, taken as the integral of $|\tilde{M}_T|^2 / 2 \tilde{n}_{-1,\mathrm{c}}$ over the condensate area, is consistent with
the magnon condensate numbers determined from momentum-space measurements \footnote{The fraction of all trapped atoms within the magnon condensate, determined from the transverse
magnetization, varies from 1.8\% to 3.7\%.  For identically prepared samples, from measurements of the magnon momentum-space distribution, we estimate this fraction to be 2.6\%.
Shot-to-shot variations in the former measurements are consistent with fluctuations in the spin rotation pulses that pump a variable number of magnons initially into the gas.}.

This transverse magnetization is inhomogeneous, both in magnitude and in phase, so that the condensed magnon gas is more aptly described as a quasi-condensate.  The spatial Fourier power spectra
of such images (Fig.\ \ref{fig:vortex}b) are concentrated in a narrow and nearly isotropic ring with spatial wavenumber $k_r$ \footnote{By imaging deliberately prepared helical spin
textures of varying wavevector, we confirm that $k_r$ lies within the resolution of our spin-sensitive imaging system; see Supplemental Material at [URL will be inserted by publisher]
for details.}.  The indicated order-parameter domain size $\xi = k_r / (2 \pi) \sim 17 \, \mu\mbox{m}$ is similar to that observed in previous studies of non-equilibrium $^{87}$Rb
spinor Bose-Einstein condensates \cite{sadl06symm,veng08helix,guzm11}.  These domains, and the associated ring in Fourier space, are observed only for magnon numbers above
the critical number; images for samples with fewer magnons show no significant transverse magnetization.

The images of the magnon quasi-condensate include singularities, regions encircled by a path along which the amplitude of the transverse magnetization is non-zero while the magnon condensate
phase $\varphi$ winds around by $\pm 2 \pi$.  We observe several such singularities in each repetition of the experiment, with the sign of the phase winding varying randomly.  We
identify these features as Mermin-Ho spin textures \cite{merm76,ho98,ohmi98}, in which the orientation of the superfluid magnetization spans a small cap about the longitudinal axis.

The inhomogeneity of the magnon quasi-condensate may be the result of the gas having been quenched rapidly across the magnon condensation transition.  Since this condensation takes
place in a uniform effective potential, it occurs simultaneously throughout the volume of the $m_F = -1$ condensate as the temperature is lowered.  The characteristic domain size $\xi$
in our measurements does not agree directly with the Kibble-Zurek theory \cite{kibb76,zure85cosmo} in two regards.  First, the Kibble-Zurek theory predicts a domain size on the order of
the thermal deBroglie wavelength $\lambda_\mathrm{dB} = (2 \pi \hbar^2 / m k_B \Tf)^{1/2} = 1.1 \, \mu\mbox{m}$, which is far smaller than observed.  Second, we imaged magnon
condensates produced with variable ramp times $t_\mathrm{ramp}$, tuning the quench time $t_Q = T_c / (dT/dt)$ between 3 and 18 s, and observed no significant change in the
characteristic domain size.  In contrast, Kibble-Zurek theory predicts a variation by a factor of 1.8 \cite{hohe77,navo15kz,chom15coherence}.

The discrepancy with Kibble-Zurek theory can be explained by the coarsening of the magnon condensate order parameter following a quench.  By varying $t_\mathrm{hold}$ between 3.5 and 11
s with a fixed $t_\mathrm{ramp} = 5 \, \mbox{s}$, we observe only slight coarsening that increases $\xi$ from 16 to 22 $\mu$m, similar to observations made in Ref.\ \cite{guzm11}.  However, it may
be that the magnon condensate is initially disordered in three dimensions, with characteristic domain size on the order of $\lambda_\mathrm{dB}$, and then rapidly coarsens until the
correlation length becomes a few times the length of the superfluid gas in its narrowest (vertical) dimension, at which point our images, which are column-integrated and limited to an
imaging resolution of around 7 $\mu$m, can finally reveal significant transverse magnetization; for this length, we may take the condensate Thomas-Fermi diameter of 6 $\mu$m at the
low-temperature trap setting. Upon the dimensional crossover to a two-dimensional system, coarsening dynamics might slow significantly.

Magnon condensation in a spinor gas offers a new system in which to study the Bose-Einstein condensation of quasi-particle excitations and of quantum gases in a uniform potential.  Our measurements of the critical magnon number, density, and momentum distribution agree with predictions based on equilibrium thermodynamics of particles held in a flat-bottomed effective potential.  However, detailed measurements of the non-equilibrium dynamics of the magnon quasi-condensate present challenges for further study.  Measurements of the real-time evolution of the gas magnetization, using repeated non-destructive spin-sensitive imaging \cite{higb05larmor,mart14magnon} with finer spatial resolution than achieved here, may address these challenges.

We acknowledge G. Edward Marti for useful discussions, Eric Copenhaver for assistance in improving the experimental apparatus, Thomas A. Mittiga for assistance during the experiment. We acknowledge the primary research support from NSF and from AFOSR through the MURI program, and secondary support for personnel from DTRA and NASA. H. K. acknowledges support by the `Studienstiftung des deutschen Volkes'.

\bibliography{allrefs_x2}

\begin{thebibliography}{29}%
\makeatletter
\providecommand \@ifxundefined [1]{%
 \@ifx{#1\undefined}
}%
\providecommand \@ifnum [1]{%
 \ifnum #1\expandafter \@firstoftwo
 \else \expandafter \@secondoftwo
 \fi
}%
\providecommand \@ifx [1]{%
 \ifx #1\expandafter \@firstoftwo
 \else \expandafter \@secondoftwo
 \fi
}%
\providecommand \natexlab [1]{#1}%
\providecommand \enquote  [1]{``#1''}%
\providecommand \bibnamefont  [1]{#1}%
\providecommand \bibfnamefont [1]{#1}%
\providecommand \citenamefont [1]{#1}%
\providecommand \href@noop [0]{\@secondoftwo}%
\providecommand \href [0]{\begingroup \@sanitize@url \@href}%
\providecommand \@href[1]{\@@startlink{#1}\@@href}%
\providecommand \@@href[1]{\endgroup#1\@@endlink}%
\providecommand \@sanitize@url [0]{\catcode `\\12\catcode `\$12\catcode
  `\&12\catcode `\#12\catcode `\^12\catcode `\_12\catcode `\%12\relax}%
\providecommand \@@startlink[1]{}%
\providecommand \@@endlink[0]{}%
\providecommand \url  [0]{\begingroup\@sanitize@url \@url }%
\providecommand \@url [1]{\endgroup\@href {#1}{\urlprefix }}%
\providecommand \urlprefix  [0]{URL }%
\providecommand \Eprint [0]{\href }%
\providecommand \doibase [0]{http://dx.doi.org/}%
\providecommand \selectlanguage [0]{\@gobble}%
\providecommand \bibinfo  [0]{\@secondoftwo}%
\providecommand \bibfield  [0]{\@secondoftwo}%
\providecommand \translation [1]{[#1]}%
\providecommand \BibitemOpen [0]{}%
\providecommand \bibitemStop [0]{}%
\providecommand \bibitemNoStop [0]{.\EOS\space}%
\providecommand \EOS [0]{\spacefactor3000\relax}%
\providecommand \BibitemShut  [1]{\csname bibitem#1\endcsname}%
\let\auto@bib@innerbib\@empty
\bibitem [{\citenamefont {Volovik}(2008)}]{volo0820years}%
  \BibitemOpen
  \bibfield  {author} {\bibinfo {author} {\bibfnamefont {G.~E.}\ \bibnamefont
  {Volovik}},\ }\href@noop {} {\bibfield  {journal} {\bibinfo  {journal}
  {Journal of Low Temperature Physics}\ }\textbf {\bibinfo {volume} {153}},\
  \bibinfo {pages} {266} (\bibinfo {year} {2008})}\BibitemShut {NoStop}%
\bibitem [{\citenamefont {Demokritov}\ \emph {et~al.}(2006)\citenamefont
  {Demokritov}, \citenamefont {Demidov}, \citenamefont {Dzyapko}, \citenamefont
  {Melkov}, \citenamefont {Serga}, \citenamefont {Hillebrands},\ and\
  \citenamefont {Slavin}}]{demo06magnonpumping}%
  \BibitemOpen
  \bibfield  {author} {\bibinfo {author} {\bibfnamefont {S.~O.}\ \bibnamefont
  {Demokritov}}, \bibinfo {author} {\bibfnamefont {V.~E.}\ \bibnamefont
  {Demidov}}, \bibinfo {author} {\bibfnamefont {O.}~\bibnamefont {Dzyapko}},
  \bibinfo {author} {\bibfnamefont {G.~A.}\ \bibnamefont {Melkov}}, \bibinfo
  {author} {\bibfnamefont {A.~A.}\ \bibnamefont {Serga}}, \bibinfo {author}
  {\bibfnamefont {B.}~\bibnamefont {Hillebrands}}, \ and\ \bibinfo {author}
  {\bibfnamefont {A.~N.}\ \bibnamefont {Slavin}},\ }\href@noop {} {\bibfield
  {journal} {\bibinfo  {journal} {Nature}\ }\textbf {\bibinfo {volume} {443}},\
  \bibinfo {pages} {430} (\bibinfo {year} {2006})}\BibitemShut {NoStop}%
\bibitem [{\citenamefont {Vainio}\ \emph {et~al.}(2015)\citenamefont {Vainio},
  \citenamefont {Ahokas}, \citenamefont {Jarvinen}, \citenamefont {Lehtonen},
  \citenamefont {Novotny}, \citenamefont {Sheludiakov}, \citenamefont
  {Suominen}, \citenamefont {Vasiliev}, \citenamefont {Zvezdov}, \citenamefont
  {Khmelenko},\ and\ \citenamefont {Lee}}]{vain15magnon}%
  \BibitemOpen
  \bibfield  {author} {\bibinfo {author} {\bibfnamefont {O.}~\bibnamefont
  {Vainio}}, \bibinfo {author} {\bibfnamefont {J.}~\bibnamefont {Ahokas}},
  \bibinfo {author} {\bibfnamefont {J.}~\bibnamefont {Jarvinen}}, \bibinfo
  {author} {\bibfnamefont {L.}~\bibnamefont {Lehtonen}}, \bibinfo {author}
  {\bibfnamefont {S.}~\bibnamefont {Novotny}}, \bibinfo {author} {\bibfnamefont
  {S.}~\bibnamefont {Sheludiakov}}, \bibinfo {author} {\bibfnamefont {K.~A.}\
  \bibnamefont {Suominen}}, \bibinfo {author} {\bibfnamefont {S.}~\bibnamefont
  {Vasiliev}}, \bibinfo {author} {\bibfnamefont {D.}~\bibnamefont {Zvezdov}},
  \bibinfo {author} {\bibfnamefont {V.~V.}\ \bibnamefont {Khmelenko}}, \ and\
  \bibinfo {author} {\bibfnamefont {D.~M.}\ \bibnamefont {Lee}},\ }\href@noop
  {} {\bibfield  {journal} {\bibinfo  {journal} {Phys. Rev. Lett.}\ }\textbf
  {\bibinfo {volume} {114}} (\bibinfo {year} {2015})}\BibitemShut {NoStop}%
\bibitem [{\citenamefont {Giamarchi}\ and\ \citenamefont
  {Tsvelik}(1999)}]{giam99ladder}%
  \BibitemOpen
  \bibfield  {author} {\bibinfo {author} {\bibfnamefont {T.}~\bibnamefont
  {Giamarchi}}\ and\ \bibinfo {author} {\bibfnamefont {A.~M.}\ \bibnamefont
  {Tsvelik}},\ }\href@noop {} {\bibfield  {journal} {\bibinfo  {journal} {Phys.
  Rev. B}\ }\textbf {\bibinfo {volume} {59}},\ \bibinfo {pages} {11398}
  (\bibinfo {year} {1999})}\BibitemShut {NoStop}%
\bibitem [{\citenamefont {Nikuni}\ \emph {et~al.}(2000)\citenamefont {Nikuni},
  \citenamefont {Oshikawa}, \citenamefont {Oosawa},\ and\ \citenamefont
  {Tanaka}}]{niku00magnon}%
  \BibitemOpen
  \bibfield  {author} {\bibinfo {author} {\bibfnamefont {T.}~\bibnamefont
  {Nikuni}}, \bibinfo {author} {\bibfnamefont {M.}~\bibnamefont {Oshikawa}},
  \bibinfo {author} {\bibfnamefont {A.}~\bibnamefont {Oosawa}}, \ and\ \bibinfo
  {author} {\bibfnamefont {H.}~\bibnamefont {Tanaka}},\ }\href@noop {}
  {\bibfield  {journal} {\bibinfo  {journal} {Phys. Rev. Lett.}\ }\textbf
  {\bibinfo {volume} {84}},\ \bibinfo {pages} {5868} (\bibinfo {year}
  {2000})}\BibitemShut {NoStop}%
\bibitem [{\citenamefont {Sebastian}\ \emph {et~al.}(2006)\citenamefont
  {Sebastian}, \citenamefont {Harrison}, \citenamefont {Batista}, \citenamefont
  {Balicas}, \citenamefont {Jaime}, \citenamefont {Sharma}, \citenamefont
  {Kawashima},\ and\ \citenamefont {Fisher}}]{seba06reduction}%
  \BibitemOpen
  \bibfield  {author} {\bibinfo {author} {\bibfnamefont {S.~E.}\ \bibnamefont
  {Sebastian}}, \bibinfo {author} {\bibfnamefont {N.}~\bibnamefont {Harrison}},
  \bibinfo {author} {\bibfnamefont {C.~D.}\ \bibnamefont {Batista}}, \bibinfo
  {author} {\bibfnamefont {L.}~\bibnamefont {Balicas}}, \bibinfo {author}
  {\bibfnamefont {M.}~\bibnamefont {Jaime}}, \bibinfo {author} {\bibfnamefont
  {P.~A.}\ \bibnamefont {Sharma}}, \bibinfo {author} {\bibfnamefont
  {N.}~\bibnamefont {Kawashima}}, \ and\ \bibinfo {author} {\bibfnamefont
  {I.~R.}\ \bibnamefont {Fisher}},\ }\href@noop {} {\bibfield  {journal}
  {\bibinfo  {journal} {Nature}\ }\textbf {\bibinfo {volume} {441}},\ \bibinfo
  {pages} {617} (\bibinfo {year} {2006})}\BibitemShut {NoStop}%
\bibitem [{\citenamefont {Demidov}\ \emph {et~al.}(2008)\citenamefont
  {Demidov}, \citenamefont {Dzyapko}, \citenamefont {Demokritov}, \citenamefont
  {Melkov},\ and\ \citenamefont {Slavin}}]{demi08spont}%
  \BibitemOpen
  \bibfield  {author} {\bibinfo {author} {\bibfnamefont {V.~E.}\ \bibnamefont
  {Demidov}}, \bibinfo {author} {\bibfnamefont {O.}~\bibnamefont {Dzyapko}},
  \bibinfo {author} {\bibfnamefont {S.~O.}\ \bibnamefont {Demokritov}},
  \bibinfo {author} {\bibfnamefont {G.~A.}\ \bibnamefont {Melkov}}, \ and\
  \bibinfo {author} {\bibfnamefont {A.~N.}\ \bibnamefont {Slavin}},\
  }\href@noop {} {\bibfield  {journal} {\bibinfo  {journal} {Phys. Rev. Lett.}\
  }\textbf {\bibinfo {volume} {100}} (\bibinfo {year} {2008})}\BibitemShut
  {NoStop}%
\bibitem [{\citenamefont {Stamper-Kurn}\ and\ \citenamefont
  {Ueda}(2013)}]{stam13rmp}%
  \BibitemOpen
  \bibfield  {author} {\bibinfo {author} {\bibfnamefont {D.~M.}\ \bibnamefont
  {Stamper-Kurn}}\ and\ \bibinfo {author} {\bibfnamefont {M.}~\bibnamefont
  {Ueda}},\ }\href@noop {} {\bibfield  {journal} {\bibinfo  {journal} {Rev.
  Mod. Phys.}\ }\textbf {\bibinfo {volume} {85}},\ \bibinfo {pages} {1191}
  (\bibinfo {year} {2013})}\BibitemShut {NoStop}%
\bibitem [{\citenamefont {Marti}\ \emph {et~al.}(2014)\citenamefont {Marti},
  \citenamefont {MacRae}, \citenamefont {Olf}, \citenamefont {Lourette},
  \citenamefont {Fang},\ and\ \citenamefont {Stamper-Kurn}}]{mart14magnon}%
  \BibitemOpen
  \bibfield  {author} {\bibinfo {author} {\bibfnamefont {G.~E.}\ \bibnamefont
  {Marti}}, \bibinfo {author} {\bibfnamefont {A.}~\bibnamefont {MacRae}},
  \bibinfo {author} {\bibfnamefont {R.}~\bibnamefont {Olf}}, \bibinfo {author}
  {\bibfnamefont {S.}~\bibnamefont {Lourette}}, \bibinfo {author}
  {\bibfnamefont {F.}~\bibnamefont {Fang}}, \ and\ \bibinfo {author}
  {\bibfnamefont {D.~M.}\ \bibnamefont {Stamper-Kurn}},\ }\href@noop {}
  {\bibfield  {journal} {\bibinfo  {journal} {Phys. Rev. Lett.}\ }\textbf
  {\bibinfo {volume} {113}},\ \bibinfo {pages} {155302} (\bibinfo {year}
  {2014})}\BibitemShut {NoStop}%
\bibitem [{\citenamefont {Phuc}\ \emph {et~al.}(2013)\citenamefont {Phuc},
  \citenamefont {Kawaguchi},\ and\ \citenamefont {Ueda}}]{phuc13}%
  \BibitemOpen
  \bibfield  {author} {\bibinfo {author} {\bibfnamefont {N.~T.}\ \bibnamefont
  {Phuc}}, \bibinfo {author} {\bibfnamefont {Y.}~\bibnamefont {Kawaguchi}}, \
  and\ \bibinfo {author} {\bibfnamefont {M.}~\bibnamefont {Ueda}},\ }\href@noop
  {} {\bibfield  {journal} {\bibinfo  {journal} {Annals of Physics}\ }\textbf
  {\bibinfo {volume} {328}},\ \bibinfo {pages} {158} (\bibinfo {year}
  {2013})}\BibitemShut {NoStop}%
\bibitem [{\citenamefont {Gaunt}\ \emph {et~al.}(2013)\citenamefont {Gaunt},
  \citenamefont {Schmidutz}, \citenamefont {Gotlibovych}, \citenamefont
  {Smith},\ and\ \citenamefont {Hadzibabic}}]{gaun13flat}%
  \BibitemOpen
  \bibfield  {author} {\bibinfo {author} {\bibfnamefont {A.~L.}\ \bibnamefont
  {Gaunt}}, \bibinfo {author} {\bibfnamefont {T.~F.}\ \bibnamefont
  {Schmidutz}}, \bibinfo {author} {\bibfnamefont {I.}~\bibnamefont
  {Gotlibovych}}, \bibinfo {author} {\bibfnamefont {R.~P.}\ \bibnamefont
  {Smith}}, \ and\ \bibinfo {author} {\bibfnamefont {Z.}~\bibnamefont
  {Hadzibabic}},\ }\href@noop {} {\bibfield  {journal} {\bibinfo  {journal}
  {Phys. Rev. Lett.}\ }\textbf {\bibinfo {volume} {110}},\ \bibinfo {pages}
  {200406} (\bibinfo {year} {2013})}\BibitemShut {NoStop}%
\bibitem [{\citenamefont {Erhard}\ \emph {et~al.}(2004)\citenamefont {Erhard},
  \citenamefont {Schmaljohann}, \citenamefont {Kronjager}, \citenamefont
  {Bongs},\ and\ \citenamefont {Sengstock}}]{erha04constant}%
  \BibitemOpen
  \bibfield  {author} {\bibinfo {author} {\bibfnamefont {M.}~\bibnamefont
  {Erhard}}, \bibinfo {author} {\bibfnamefont {H.}~\bibnamefont
  {Schmaljohann}}, \bibinfo {author} {\bibfnamefont {J.}~\bibnamefont
  {Kronjager}}, \bibinfo {author} {\bibfnamefont {K.}~\bibnamefont {Bongs}}, \
  and\ \bibinfo {author} {\bibfnamefont {K.}~\bibnamefont {Sengstock}},\
  }\href@noop {} {\bibfield  {journal} {\bibinfo  {journal} {Phys. Rev. A}\
  }\textbf {\bibinfo {volume} {70}} (\bibinfo {year} {2004})}\BibitemShut
  {NoStop}%
\bibitem [{\citenamefont {Olf}\ \emph {et~al.}(2015)\citenamefont {Olf},
  \citenamefont {Fang}, \citenamefont {Marti}, \citenamefont {MacRae},\ and\
  \citenamefont {Stamper-Kurn}}]{olf15cooling}%
  \BibitemOpen
  \bibfield  {author} {\bibinfo {author} {\bibfnamefont {R.}~\bibnamefont
  {Olf}}, \bibinfo {author} {\bibfnamefont {F.}~\bibnamefont {Fang}}, \bibinfo
  {author} {\bibfnamefont {G.~E.}\ \bibnamefont {Marti}}, \bibinfo {author}
  {\bibfnamefont {A.}~\bibnamefont {MacRae}}, \ and\ \bibinfo {author}
  {\bibfnamefont {D.~M.}\ \bibnamefont {Stamper-Kurn}},\ }\href@noop {}
  {\bibfield  {journal} {\bibinfo  {journal} {Nat Phys}\ }\textbf {\bibinfo
  {volume} {11}},\ \bibinfo {pages} {720} (\bibinfo {year} {2015})}\BibitemShut
  {NoStop}%
\bibitem [{Note1()}]{Note1}%
  \BibitemOpen
  \bibinfo {note} {Here we neglect effects of non-zero temperature,
  magnon-magnon interactions, and spin-dependent potentials including the
  linear Zeeman energy of the uniform magnetic field, which, neglecting dipolar
  interactions, can be gauged away by treating the gas in a rotating
  frame}\BibitemShut {NoStop}%
\bibitem [{Note2()}]{Note2}%
  \BibitemOpen
  \bibinfo {note} {See Supplemental Material at [URL will be inserted by
  publisher] for predicted critical magnon number in each
  potential.}\BibitemShut {Stop}%
\bibitem [{\citenamefont {Guzman}\ \emph {et~al.}(2011)\citenamefont {Guzman},
  \citenamefont {Jo}, \citenamefont {Wenz}, \citenamefont {Murch},
  \citenamefont {Thomas},\ and\ \citenamefont {Stamper-Kurn}}]{guzm11}%
  \BibitemOpen
  \bibfield  {author} {\bibinfo {author} {\bibfnamefont {J.}~\bibnamefont
  {Guzman}}, \bibinfo {author} {\bibfnamefont {G.~B.}\ \bibnamefont {Jo}},
  \bibinfo {author} {\bibfnamefont {A.~N.}\ \bibnamefont {Wenz}}, \bibinfo
  {author} {\bibfnamefont {K.~W.}\ \bibnamefont {Murch}}, \bibinfo {author}
  {\bibfnamefont {C.~K.}\ \bibnamefont {Thomas}}, \ and\ \bibinfo {author}
  {\bibfnamefont {D.~M.}\ \bibnamefont {Stamper-Kurn}},\ }\href@noop {}
  {\bibfield  {journal} {\bibinfo  {journal} {Phys. Rev. A}\ }\textbf {\bibinfo
  {volume} {84}},\ \bibinfo {pages} {063625} (\bibinfo {year}
  {2011})}\BibitemShut {NoStop}%
\bibitem [{Note3()}]{Note3}%
  \BibitemOpen
  \bibinfo {note} {The fraction of all trapped atoms within the magnon
  condensate, determined from the transverse magnetization, varies from 1.8\%
  to 3.7\%. For identically prepared samples, from measurements of the magnon
  momentum-space distribution, we estimate this fraction to be 2.6\%.
  Shot-to-shot variations in the former measurements are consistent with
  fluctuations in the spin rotation pulses that pump a variable number of
  magnons initially into the gas.}\BibitemShut {Stop}%
\bibitem [{Note4()}]{Note4}%
  \BibitemOpen
  \bibinfo {note} {By imaging deliberately prepared helical spin textures of
  varying wavevector, we confirm that $k_r$ lies within the resolution of our
  spin-sensitive imaging system; see Supplemental Material at [URL will be
  inserted by publisher] for details.}\BibitemShut {Stop}%
\bibitem [{\citenamefont {Sadler}\ \emph {et~al.}(2006)\citenamefont {Sadler},
  \citenamefont {Higbie}, \citenamefont {Leslie}, \citenamefont
  {Vengalattore},\ and\ \citenamefont {Stamper-Kurn}}]{sadl06symm}%
  \BibitemOpen
  \bibfield  {author} {\bibinfo {author} {\bibfnamefont {L.}~\bibnamefont
  {Sadler}}, \bibinfo {author} {\bibfnamefont {J.}~\bibnamefont {Higbie}},
  \bibinfo {author} {\bibfnamefont {S.}~\bibnamefont {Leslie}}, \bibinfo
  {author} {\bibfnamefont {M.}~\bibnamefont {Vengalattore}}, \ and\ \bibinfo
  {author} {\bibfnamefont {D.}~\bibnamefont {Stamper-Kurn}},\ }\href@noop {}
  {\bibfield  {journal} {\bibinfo  {journal} {Nature}\ }\textbf {\bibinfo
  {volume} {443}},\ \bibinfo {pages} {312} (\bibinfo {year}
  {2006})}\BibitemShut {NoStop}%
\bibitem [{\citenamefont {Vengalattore}\ \emph {et~al.}(2008)\citenamefont
  {Vengalattore}, \citenamefont {Leslie}, \citenamefont {Guzman},\ and\
  \citenamefont {Stamper-Kurn}}]{veng08helix}%
  \BibitemOpen
  \bibfield  {author} {\bibinfo {author} {\bibfnamefont {M.}~\bibnamefont
  {Vengalattore}}, \bibinfo {author} {\bibfnamefont {S.}~\bibnamefont
  {Leslie}}, \bibinfo {author} {\bibfnamefont {J.}~\bibnamefont {Guzman}}, \
  and\ \bibinfo {author} {\bibfnamefont {D.}~\bibnamefont {Stamper-Kurn}},\
  }\href@noop {} {\bibfield  {journal} {\bibinfo  {journal} {Phys. Rev. Lett.}\
  }\textbf {\bibinfo {volume} {100}},\ \bibinfo {pages} {170403} (\bibinfo
  {year} {2008})}\BibitemShut {NoStop}%
\bibitem [{\citenamefont {Mermin}\ and\ \citenamefont {Ho}(1976)}]{merm76}%
  \BibitemOpen
  \bibfield  {author} {\bibinfo {author} {\bibfnamefont {N.~D.}\ \bibnamefont
  {Mermin}}\ and\ \bibinfo {author} {\bibfnamefont {T.-L.}\ \bibnamefont
  {Ho}},\ }\href@noop {} {\bibfield  {journal} {\bibinfo  {journal} {Phys. Rev.
  Lett.}\ }\textbf {\bibinfo {volume} {36}},\ \bibinfo {pages} {594} (\bibinfo
  {year} {1976})}\BibitemShut {NoStop}%
\bibitem [{\citenamefont {Ho}(1998)}]{ho98}%
  \BibitemOpen
  \bibfield  {author} {\bibinfo {author} {\bibfnamefont {T.-L.}\ \bibnamefont
  {Ho}},\ }\href@noop {} {\bibfield  {journal} {\bibinfo  {journal} {Phys. Rev.
  Lett.}\ }\textbf {\bibinfo {volume} {81}},\ \bibinfo {pages} {742} (\bibinfo
  {year} {1998})}\BibitemShut {NoStop}%
\bibitem [{\citenamefont {Ohmi}\ and\ \citenamefont {Machida}(1998)}]{ohmi98}%
  \BibitemOpen
  \bibfield  {author} {\bibinfo {author} {\bibfnamefont {T.}~\bibnamefont
  {Ohmi}}\ and\ \bibinfo {author} {\bibfnamefont {K.}~\bibnamefont {Machida}},\
  }\href@noop {} {\bibfield  {journal} {\bibinfo  {journal} {J. Phys. Soc.
  Jpn.}\ }\textbf {\bibinfo {volume} {67}},\ \bibinfo {pages} {1822} (\bibinfo
  {year} {1998})}\BibitemShut {NoStop}%
\bibitem [{\citenamefont {Kibble}(1976)}]{kibb76}%
  \BibitemOpen
  \bibfield  {author} {\bibinfo {author} {\bibfnamefont {T.~W.~B.}\
  \bibnamefont {Kibble}},\ }\href@noop {} {\bibfield  {journal} {\bibinfo
  {journal} {J. Phys. A}\ }\textbf {\bibinfo {volume} {9}},\ \bibinfo {pages}
  {1387} (\bibinfo {year} {1976})}\BibitemShut {NoStop}%
\bibitem [{\citenamefont {Zurek}(1985)}]{zure85cosmo}%
  \BibitemOpen
  \bibfield  {author} {\bibinfo {author} {\bibfnamefont {W.~H.}\ \bibnamefont
  {Zurek}},\ }\href@noop {} {\bibfield  {journal} {\bibinfo  {journal}
  {Nature}\ }\textbf {\bibinfo {volume} {317}},\ \bibinfo {pages} {505}
  (\bibinfo {year} {1985})}\BibitemShut {NoStop}%
\bibitem [{\citenamefont {Hohenberg}\ and\ \citenamefont
  {Halperin}(1977)}]{hohe77}%
  \BibitemOpen
  \bibfield  {author} {\bibinfo {author} {\bibfnamefont {P.~C.}\ \bibnamefont
  {Hohenberg}}\ and\ \bibinfo {author} {\bibfnamefont {B.~I.}\ \bibnamefont
  {Halperin}},\ }\href@noop {} {\bibfield  {journal} {\bibinfo  {journal} {Rev.
  Mod. Phys.}\ }\textbf {\bibinfo {volume} {49}},\ \bibinfo {pages} {435}
  (\bibinfo {year} {1977})}\BibitemShut {NoStop}%
\bibitem [{\citenamefont {Navon}\ \emph {et~al.}(2015)\citenamefont {Navon},
  \citenamefont {Gaunt}, \citenamefont {Smith},\ and\ \citenamefont
  {Hadzibabic}}]{navo15kz}%
  \BibitemOpen
  \bibfield  {author} {\bibinfo {author} {\bibfnamefont {N.}~\bibnamefont
  {Navon}}, \bibinfo {author} {\bibfnamefont {A.~L.}\ \bibnamefont {Gaunt}},
  \bibinfo {author} {\bibfnamefont {R.~P.}\ \bibnamefont {Smith}}, \ and\
  \bibinfo {author} {\bibfnamefont {Z.}~\bibnamefont {Hadzibabic}},\
  }\href@noop {} {\bibfield  {journal} {\bibinfo  {journal} {Science}\ }\textbf
  {\bibinfo {volume} {347}},\ \bibinfo {pages} {167} (\bibinfo {year}
  {2015})}\BibitemShut {NoStop}%
\bibitem [{\citenamefont {Chomaz}\ \emph {et~al.}(2015)\citenamefont {Chomaz},
  \citenamefont {Corman}, \citenamefont {Bienaimé}, \citenamefont
  {Desbuquois}, \citenamefont {Weitenberg}, \citenamefont {Nascimbène},
  \citenamefont {Beugnon},\ and\ \citenamefont {Dalibard}}]{chom15coherence}%
  \BibitemOpen
  \bibfield  {author} {\bibinfo {author} {\bibfnamefont {L.}~\bibnamefont
  {Chomaz}}, \bibinfo {author} {\bibfnamefont {L.}~\bibnamefont {Corman}},
  \bibinfo {author} {\bibfnamefont {T.}~\bibnamefont {Bienaimé}}, \bibinfo
  {author} {\bibfnamefont {R.}~\bibnamefont {Desbuquois}}, \bibinfo {author}
  {\bibfnamefont {C.}~\bibnamefont {Weitenberg}}, \bibinfo {author}
  {\bibfnamefont {S.}~\bibnamefont {Nascimbène}}, \bibinfo {author}
  {\bibfnamefont {J.}~\bibnamefont {Beugnon}}, \ and\ \bibinfo {author}
  {\bibfnamefont {J.}~\bibnamefont {Dalibard}},\ }\href@noop {} {\bibfield
  {journal} {\bibinfo  {journal} {Nat Commun}\ }\textbf {\bibinfo {volume} {6}}
  (\bibinfo {year} {2015})}\BibitemShut {NoStop}%
\bibitem [{\citenamefont {Higbie}\ \emph {et~al.}(2005)\citenamefont {Higbie},
  \citenamefont {Sadler}, \citenamefont {Inouye}, \citenamefont {Chikkatur},
  \citenamefont {Leslie}, \citenamefont {Moore}, \citenamefont {Savalli},\ and\
  \citenamefont {Stamper-Kurn}}]{higb05larmor}%
  \BibitemOpen
  \bibfield  {author} {\bibinfo {author} {\bibfnamefont {J.}~\bibnamefont
  {Higbie}}, \bibinfo {author} {\bibfnamefont {L.}~\bibnamefont {Sadler}},
  \bibinfo {author} {\bibfnamefont {S.}~\bibnamefont {Inouye}}, \bibinfo
  {author} {\bibfnamefont {A.~P.}\ \bibnamefont {Chikkatur}}, \bibinfo {author}
  {\bibfnamefont {S.~R.}\ \bibnamefont {Leslie}}, \bibinfo {author}
  {\bibfnamefont {K.~L.}\ \bibnamefont {Moore}}, \bibinfo {author}
  {\bibfnamefont {V.}~\bibnamefont {Savalli}}, \ and\ \bibinfo {author}
  {\bibfnamefont {D.~M.}\ \bibnamefont {Stamper-Kurn}},\ }\href@noop {}
  {\bibfield  {journal} {\bibinfo  {journal} {Phys. Rev. Lett.}\ }\textbf
  {\bibinfo {volume} {95}},\ \bibinfo {pages} {050401} (\bibinfo {year}
  {2005})}\BibitemShut {NoStop}%
\end{thebibliography}%

\end{document}